\documentclass[doublecol]{epl2} % for 2 columns style with line numbers
\usepackage{dsfont}
\usepackage{graphicx}
\usepackage{epstopdf} 
\usepackage{subfig}
\usepackage{amsmath}
\usepackage{amssymb}

\DeclareMathOperator*{\argmax}{arg\,max}

\title{Information inefficiency in a random linear economy model}
\shorttitle{Information inefficiency in a linear economy model} 

\author{Jo\~ao Pedro Jeric\'o\inst{1} \and Renato Vicente\inst{2}}

\institute{                    
  \inst{1} Dep. de F{\'\i}sica Geral, Instituto de F{\'\i}sica, Universidade de S\~ao Paulo, CP 66319, 05315-970, S\~ao Paulo-SP, Brazil\\
  \inst{2} Dept. of Applied Mathematics, Instituto de Matem\'atica e Estat{\'\i}stica, Universidade de S\~ao Paulo, 05508-090,S\~ao Paulo-SP, Brazil 
}

\abstract{We study the effects of introducing information inefficiency
  in a model for a random linear economy with a representative consumer. This is done  by considering  statistical, instead of classical, economic general equilibria. Employing two different approaches we show that inefficiency increases the consumption set of a consumer but decreases her expected utility. In this scenario economic activity grows  while  welfare shrinks, what is contrary to the behavior obtained by considering a rational consumer.}

\begin{document}

\maketitle

\section{Introduction}

Standard microeconomic theory states that an agent faced with a choice  of goods or services to purchase will maximize an
expected utility function given her constraints \cite{MasColell}. It also predicts an equilibrium state resulting from these agents simultanously and independently trying to optimize their own choices. 

Over the last decade these assumptions have been under
intense scrutiny \cite{Bouchaud08, Kirman10}. Among other factors, the
2007-2008 mortgage crisis, for which the predictions made by
mainstream theories were ineffective in preventing, and the growing
empirical body of evidence being gathered in the field of Behavioral
Economics have shown that perfect rationality may be a very poor
proxy for the observed behavior of economic agents.

In fact, it has been shown that modelling agents to have zero
intelligence also produces realistic results \cite{Gode93, Smith03}
due to general statistical properties that may emerge in interacting
systems. As argued by the authors in \cite{Smith03}, treating economic
agents as having full rationality and perfect knowledge is clearly a
significant assumption, so why not model the other end of the
spectrum?  It has been shown in \cite{Yakovenko00}
that, for an economy in which trades occur in a random fashion, the
equilibrium wealth distribution is a Gibbs distribution. This is very close to
empirical observations for several countries, at least in the bulk of the distribution. By adding a rate of return on capital,
power law distributions are obtained \cite{Bouchaud00}, what is again consistent with
empirical data for top earners.

Statistical mechanics provides a robust framework, based on maximum entropy principles, for identifying and calculating macroscopic quantities 
relevant for the  description of interacting  systems.
Within this framework, the temperature of a system represents how large  observed
deviations from the ground state are likely to be. Clearly, a zero temperature system is always at the configuration
which minimizes its energy function, while a system at infinite
temperature will be at any configuration allowed by its constraints with equal probability.

These considerations suggest a way to model consumers that may turn to be more realistic: the degree of rationality may be represented as a parameter within a statistical mechanics model with energy given by the negative utility. A similar approach has already appeared in the contexts of game theory \cite{BrockDurlauf01, Blume93}
and general equilibrium  \cite{Foley94, Foley96}.

Although deviations from maximum utility are here named, in accordance with economics literature, ``suboptimal'' or ``inefficient'', it is  common knowledge in statistical inference that a ``softmax'' decision
rule  can actually be beneficial if the agent  has incomplete knowledge about the universe of available options. 

In this paper we discuss two approaches for considering consumer irrationality and show that they qualitatively lead to the same behavior for important macroeconomic quantities.

\section{The Random Linear Economy Model}

In a random linear economy \cite{DeMartinoMarsili04} there are  $M$
goods, $N$ technologies and a single representative consumer. Let  $n = N/M$ represent the multiplicity of technologies per good.

A consumer is endowed by an initial bundle of goods $x_0 = (x_0^1, \ldots, x_0^M)$. We will be interested in an ensemble of economies with a given technology multiplicity. We suppose an endowment $x_0^\mu \geq 0$ is drawn
independently from an exponential distribution. Consumer preferences are ordered according to a separable  utility function
\begin{equation*}
 U(x) = \sum_{\mu=1}^M u(x^\mu).
\end{equation*}
The consumer  optimizes its utility subject to a budget constraint $x\cdot p \le x_0\cdot p$, where $p$ represents a vector of prices. 

In a closed economy,  markets clear at equilibrium so that firms inputs  have to come from consumer's
initial endowment. The $N$ dimensional production scale vector $s$ satisfies 
\begin{equation}
x = x_0 + \sum_{i=1}^N s_i \xi_i.
\label{eq:market_clearing}
\end{equation}

Every firm in the market is equipped with a random technology
$\xi_i = (\xi_i^1, \ldots, \xi_i^\mu)$, where $\xi_i^\mu<0$ represents
an input and $\xi_i^\mu>0$ represents an output. Thus, a firm represents a transformation in the space of goods given by the $M$ dimensional
vector $s_i \xi_i$.  The firm has to decide at what scale $s_i$ it will operate its technology so that profits $s_i p_i \cdot \xi_i$ are maximized.
We consider normally distributed elements $\xi_i^\mu$  with variance $1/M$   constrained by
$\sum_{\mu=1}^M \xi_i^\mu = -\epsilon$. For $\epsilon >0$ every technology is inefficient and it is impossible to couple two or more of them in order to produce an infinite amount of a good.

The market clearing condition implies that the utilities and profits can be maximized simultaneously. Optimal consumer bundle  $x^* = \argmax_x U(x)$ and production scales $s_i^* = \argmax_{s_i} s_i p_i \cdot \xi_i$ are attained when  prices are set
to satisfy the first order condition for the consumer utility  maximization
problem $p_\mu = \frac{\partial U}{\partial x_\mu}$. At the equilibrium,  firms production and the market prices can be
set to satisfy consumer's desired demand, and no actor in the market
has an incentive to deviate from this state.

Market clearing, however, carries a strong restriction. If we multiply both
sides of the equation \eqref{eq:market_clearing} by $p$, we get
\begin{equation}
  \label{eq:market_clearing_p}
  p\cdot (x - x_0) = \sum_i s_i p \cdot \xi_i.
\end{equation}
The budget condition implies a non-positive left side,  while the right hand side of this equation is necessarily non-negative, as the scale of production can be made $s_i = 0$ to avoid losses. The equation is, therefore, only satisfied when agents spend their budget to the limit, to say, when $p\cdot x = p \cdot x_0$,  and when firm profits vanish, $p\cdot \xi_i = 0$, or firms leave the market, and $s_i = 0$. 

Equation \eqref{eq:market_clearing_p} also implies that not more than $M$ different technologies can be simultaneously active at  equilibrium. This can be seen by observing that, for $n>1$, $p\cdot \xi_i = 0$ represents an overdetermined linear system with $N$ equations and $M$ variables.

This model has some very interesting properties which are described at
length in \cite{DeMartinoMarsili04} and more recently explored in \cite{Bardoscia15}. In particular, given an ensemble of economies, we can calculate the  probability distributions of $x$ and $s$
(and therefore of $p$) and find that all macroscopic quantities derived
from these two quantities depend on the technology multiplicity per good
$n = N/M$. When $n<2$, the market is competitive and the
fraction of active firms $\phi = \sum_i \mathbb{I}(s_i > 0) / N$ is
around $\phi = 0.5$. As each firm has on average half the goods
as inputs and half as outputs, when $n<2$, the number of  
technologies is not sufficient to produce every combination of $M$ goods demanded. The space of goods is not spanned by the technologies available.

When $n>2$, however, there are so many technologies to choose from that
 it is possible  to find  $M$ linear independent firms
that span the whole space of goods. The market thus becomes
monopolistic and $\phi$  vanishes asymptotically with $n$ as the maximum number of active technologies is $M$.

In Economics the Gross Domestic Product (GDP) is a measure of the total market value of final goods produced in a given period. In the random linear economy model, the consequence of the market clearing condition  \eqref{eq:market_clearing_p} is that no net market value is created by the transformation process. An alternative measure of economic activity within the model is provided by 
\begin{equation}
  \label{eq:5}
  Y = \frac{\sum_{\mu = 1}^M |x_\mu - x_0^\mu|p_\mu}{2 \sum_{\mu = 0}^M p_\mu}.
\end{equation}
In \cite{DeMartinoMarsili04} it is shown that in the competitive
regime $n<2$, a new technology has a positive effect
on economic activity $Y$, while in the monopolistic regime $n>2$, new technologies have
negligible impact on the economic activity. Our goals in this paper is  revisiting this result in a  setting  where the representative consumer is  an inefficient optimizer.

In what follows we do that in two different ways. In the first approach we employ a maximum entropy argument for replacing the idea that the consumer is a hyperrational maximizer. In the second approach we suppose that the consumer actually optimizes a noisy version of her utility.

\section{The Irrational Consumer}

We now turn to the problem of  treating the consumer as an
agent that makes suboptimal choices with as 
little extra assumptions for her behavior as possible. 

An answer for this type of question is provided by statistical mechanics in the form of the following inference problem: find the
probability distribution for $x$ that has maximum entropy given the
constraints that $P(x)$ is a probability distribution  and the average utility $\bar{U}$ is observed\cite{JaynesPRL}. Using the simplifying notation that $\int dx =
\int_0^\infty \cdots \int_0^\infty \prod_\mu dx_\mu$, we want to find
the function $P(x)$ that maximizes the functional

\begin{align}
  \label{eq:11}
  S[P(x)]  = & -\int_0^\infty dx P(x)\log P(x) - \alpha 
  \left(\int dx P(x) - 1\right) \nonumber \\ & + \beta \left(\int dx U(x) P(x) - \bar{U} \right),
\end{align}
where $\alpha$ and $\beta$ are Lagrange multipliers for the
constraints. The well known solution for this entropy maximization
problem is the Gibbs distribution:

\begin{equation}
  \label{eq:gibbs}
  P(x) = \frac{1}{Z(\beta)} e^{\beta U(x)},
\end{equation}

where $Z = \int_0^\infty dx e^{\beta U(x)}$ is the normalization term
and $\beta$ is a free parameter that controls how much we allow for
deviations from the maximum. It is clear that in the limit $\beta \to \infty$ the
distribution collapses to a Dirac delta in its maximum and the
inference reduces to the classical consumer problem. We, however, have found in $\beta$
 a way to introduce deviations from the maximum. Additionally, the maximum entropy argument assures we are not introducing unjustified assumptions about consumer's behavior.

Because market clearing still holds, we can plug equation
\eqref{eq:market_clearing} directly into the Gibbs distribution. The
probability of the consumer choosing a bundle $x$ becomes the probability of   firms setting a production vector $s$, given by:

\begin{equation}
  \label{eq:1}
  P(s | \xi, x_0) = \frac{1}{Z} e^{\beta U(x_0 + \sum_i s_i \xi_i)}
\end{equation}

\begin{figure}[!ht]
  \centering
  \includegraphics[width=.45\textwidth]{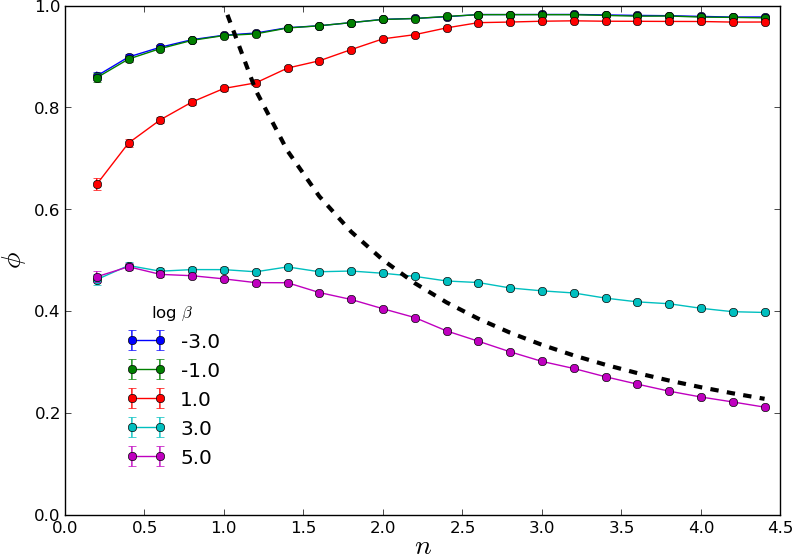}
  \caption{Fraction of active firms $\phi$ as a function of the
    technological multiplicity $n$, for different values of $\beta$. In these simulations $\epsilon=0.05$ and 
    $M=64$ averaged over $400$ runs.
    As expected, $\beta\rightarrow\infty$  implies $\phi\rightarrow 0.5$ for $n<<2$. Above the dashed line no global prices capable of supporting economic activity can be defined. }
  \label{fig:s_pos_n}
\end{figure}

Although this distribution is intractable, except for
$\beta \to \infty$, we still can find the expected values of
macroeconomic variables for a random linear economy  by  Metropolis sampling  averaged over the disorder $(\xi, x_0)$. We proceed by first finding the average fraction of active
firms $\langle \phi \rangle_{\xi, x_0}$ at the equilibrium as a
function of $\beta$. The results are shown in Figure
\ref{fig:s_pos_n}. We see that the inefficiency parameter $\beta$ has
a strong influence on the ``two regimes'' observed in the efficient case, which is shown by making $\beta > 10^5$. For lower values, the
monopolistic regime vanishes, and after a certain point the fraction of
active firms actually increases with $n$.

More interesting, however, are the behavior of other macroeconomic
variables of the model. By definition, a  rational consumer increases her utility as $n$ increases. If the supply of trades is expanded by the introduction of new firms operating in a market, the agent can always choose to keep her 
current exchange or replace it for a better one. 

However, when agents are irrational that is not not always the case. In this case we can  compute numerically
the average utility per good
$\langle u \rangle$ for the consumer as a
function of $n$:
\begin{equation}
  \label{eq:4}
  \langle u \rangle = \frac{1}{M} \langle U \rangle_{\xi, x_0} =
  \frac{1}{M} \int dx \int d\xi dx_0 U(x)
  \frac{1}{Z} e^{\beta U(x)}.
\end{equation}

\begin{figure}[!ht]
  \centering
  \includegraphics[width=.45\textwidth]{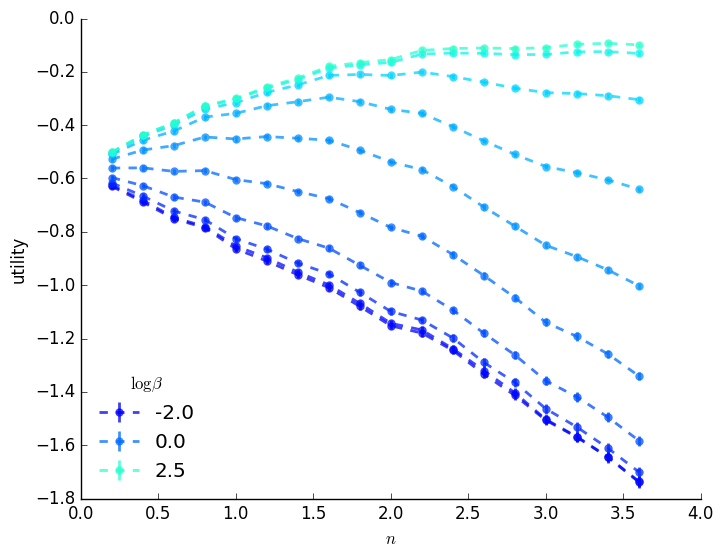}
  \includegraphics[width=.45\textwidth]{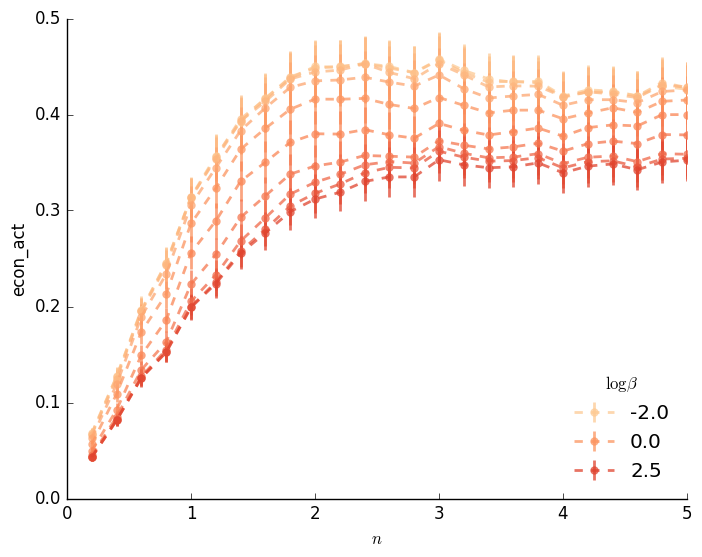}
  \caption{\textbf{Top.} Average utility of the consumer as a function of $n$, for
    different values of $\beta$. \textbf{Bottom.} Average economic activity for the
    economy as a function of $n$.}
  \label{fig:u_and_GDP}
\end{figure}

As can be seen in Figure \ref{fig:u_and_GDP}, for large $\beta$, $\langle u \rangle$ is an increasing function of $n$, as expected. However, when inefficiency is increased, the average utility of the consumer becomes a
decreasing function. This result corroborates a number of
empirical observations in Behavioral Economics claiming that consuming options can turn to be excessive \cite{Lepper00, Schwartz02}: as the number of
options available to a consumer without perfect knowledge
and full rationality increases the probability of choice mistakes also
increases.

As a proxy for economic activity, we also calculate the absolute amount of goods traded, that is given by
\begin{equation}
  \label{eq:6}
  \Lambda = \frac{1}{M} \sum_{\mu = 1}^M \left| x_\mu - x_0^\mu \right|
\end{equation}

We observe in the simulations depicted in the lower panel of  Figure \ref{fig:u_and_GDP} that
the economic activity always increases as a function of the
technological density $n$, but the rate of increase in $\Lambda$ also
increases with the inefficiency, as opposed to the utility. This is our main observation: irrational choices enhance the amount of goods
traded while worsening the quality of what is acquired, decreasing consumer's utility.

A difficulty with our analysis still remains: whereas originally
the prices were uniquely defined by the first order condition of the
consumer's optimization problem, in this new framework there is no
such restraint. Given a probability distribution $P(x | \xi, x_0)$,
it is an ill-posed problem to define the price vector from it. In
fact, any price vector $p$ satisfying
\begin{equation}
  \label{eq:3}
p \cdot \xi_i = 0, \text{ for all } i \text{ if } s_i > 0  
\end{equation}
would be an equally good choice.

Since we have $N_a$ equations
of this type, one for each active firm, and $M$ unknown variables
$p_\mu$, the above equations have no solution for $N_a > M$, provided that $\xi_i$ are linearly independent.

 In terms of $\phi$, this
means that for $\phi > 1/n$ there is no global price vector $p$ that
supports  economic activity. The  curve $\phi = 1/n$ is plotted (dashed line) in Figure \ref{fig:s_pos_n}. The model only stays, for all $n$, in the region where global prices are possible when very close to rationality ($\beta$ large). When there is a level of inefficiency, there is always
a threshold $n$, above which no global prices can exist. 

This does not mean that no
market transactions will take place, instead it means that in this approach prices must be defined for each of the $N$ suppliers. 

\section{Noisy Utility}

As an alternative along the lines of \cite{MarsiliRoudi13}, we can modify the classical setup so that
the equilibrium is still reached from a maximization problem solved by the
consumer. We suppose now that the representative consumer thinks she is rationally optimizing an utility $U$. However, she is actually optimizing another utility $U_T$, a noisy version of $U$.

Call
$U_T$ the ``true'' utility function and write it as
\begin{equation}
  \label{eq:2}
  U_T(x) = U(x) + h \cdot x,
\end{equation}
where $h$ is a $M$ dimensional vector with all entries sampled independently from an
exponential distribution with scale $\lambda$. This extra term
represents a part of the utility unknown by the consumer herself. We point out that the true utility function $U_T$ is still a
concave function and therefore the equilibrium is still unique and
well behaved.

It can be shown \cite{MarsiliRoudi13} that the resulting probability
distribution for $x$, when the consumer is maximizing the utility
distribution $U_T(x)$, is the Gibbs distribution for $U(x)$. This
allows us to reformulate the problem in terms of a consumer that does not
choose the ``optimal'' utility, keeping a
well defined price vector which results from  the usual first order
conditions. For example, for the utility
$U(x) = \sum_\mu \log(x_\mu)$ the price of a good would be given by
$p_\mu = \partial_\mu U_t(x) = \frac{1}{x_\mu} + h_\mu$.

\begin{figure}[!ht]
  \centering
  \includegraphics[width=.45\textwidth]{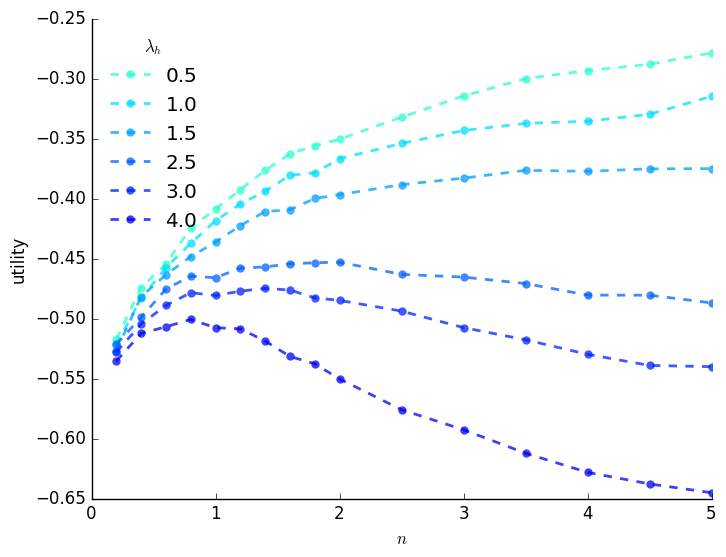}
  \includegraphics[width=.45\textwidth]{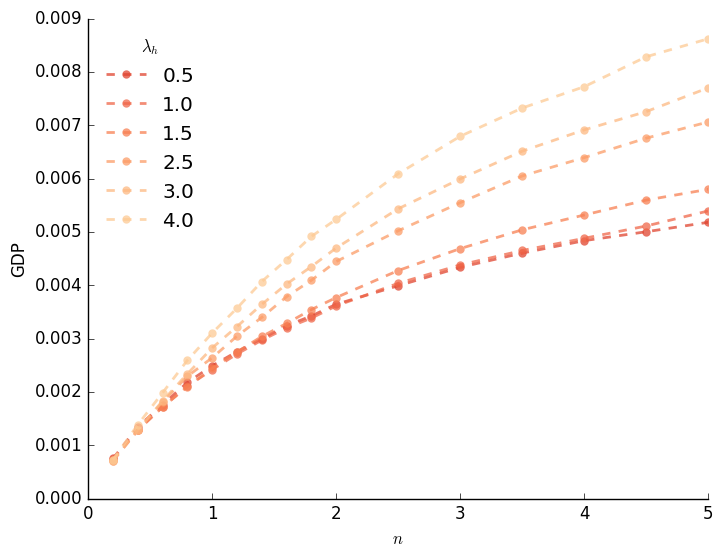}
  \caption{\textbf{Top.} Average observed utility of the consumer as a
    function of $n$ in the unoberserved utility framework, for
    different values of $\beta$. \textbf{Bottom.} Average GDP for the
    economy as a function of $n$.}
  \label{fig:u_and_GDP_stoc}
\end{figure}

The results for the fraction of active firms $\phi$ are now the same of the rational agent case and we still have a
competitive regime for $n<2$ and a monopolistic regime for $n>2$. However, the behavior of the macroeconomic quantities reproduces that of the
inefficient consumer.
Figure \ref{fig:u_and_GDP_stoc} depicts solutions of the
maximization process of $U_T$  for various values of the
scale $\lambda$ for the exponential distribution of $h_\mu$.

As in the previous case, above a threshold $\lambda$, the
observed utility decreases with $n$, while the GDP 
increases.

In this approach we had to introduce additional assumptions about consumer's utility.  As it was noticed by the authors of \cite{MarsiliRoudi13}, the Gibbs distribution for
the observed depends on very general
conditions for the unknown part,  therefore, the results we have 
obtained are supposed to be robust to different choices in the
perturbed utility.

\section{Conclusion}

  We have shown that the introduction of irrationality in the behavior 
  a representative consumer implies in macroeconomic behavior very distinct of
  what is the derived from classical rationality axioms. 
  
  Provided the coupling between demand and supply imposed by a market clearing scenario, 
  as inefficiency is introduced to the consumer behavior, the supply side is exposed to less selection and the consumer is flooded by an overabundance of low utility goods. What corroborates results from Behavioral Economics \cite{Lepper00, Schwartz02}. 
  
  With less selection the supply side produces more and GDP grows. Thus a larger number of consumption options may result in poorer choices while increasing economic
activity. 

Our conclusion may be intriguing: for a whole country's economy it is seems desirable that its consumers are less rational agents as less efficiency in
choice leads to enhanced GDP and tax revenues. If
we thinks of country development as an evolutionary process, there 
would be incentives for less rational consumers, penalizing the lower
economic activity of more a efficient demand side. Additionally, if the average utility can be considered a proxy for individual fitness, then a more rational
consumer would thrive in a country with low choice efficiency, suggesting that an interesting competition process may take place.

This is certainly too deep a conclusion to be arrived from just this
simple model. However, the results may be robust for other equilibrium
scenarios, and we hope this paper sets the stage for future research
on the topic.

\acknowledgments J.P.J. is thankful for the grant conceded by FAPESP,
process number 2012/19521-8. We also thank Matteo Marsili for his
contributions to this paper.

\end{document}